\documentclass{article}
\usepackage{graphicx}  
\usepackage{amsmath}   
\usepackage[compress]{cite}
\usepackage{amssymb}   
\usepackage{bm} 
\usepackage{dcolumn}
\usepackage{color}
\usepackage{mathrsfs}
\usepackage{amsfonts}
\usepackage{varioref}
\usepackage{authblk}
\RequirePackage[colorlinks,citecolor=blue,urlcolor=magenta,linkcolor=blue]{hyperref}
\def\gr{general relativity}
\def\RN{Reissner-Nordstr\"{o}m }

\labelformat{section}{Section #1} 
\labelformat{subsection}{Section #1} 
\labelformat{subsubsection}{Section #1}
\labelformat{subsubsubsection}{Section #1}
\labelformat{equation}{Eq.~(#1)} 
\labelformat{figure}{Fig.~#1} 
\labelformat{subfigure}{Fig.~\thefigure#1} 
\labelformat{table}{Tab.~#1} 
\labelformat{appendix}{Appendix #1}
\title{Constraining some Horndeski gravity theories}
\author{Sourav Bhattacharya$^{1,2}$ \footnote{sbhatta@iitrpr.ac.in} 
~~and~
Sumanta Chakraborty$^{1,3}$ \footnote{sumantac.physics@gmail.com}\\
$^{1}$\small{IUCAA, Post Bag 4, Ganeshkhind, Pune University Campus, Pune 411 007, India}\\
$^{2}$\small{Department of Physics, IIT Ropar, Ropar 140 001, India}\\
$^{3}$\small{Department of Theoretical Physics, Indian Association for the Cultivation of Science, Kolkata 700032, India}}
\begin{document}

\maketitle
\begin{abstract}
\noindent
We discuss two spherically symmetric solutions admitted by the Horndeski (or scalar tensor) theory in the context of solar system and astrophysical scenarios. One of these solutions is derived for Einstein-Gauss-Bonnet gravity, while the other originates from the coupling of the Gauss-Bonnet invariant with a scalar field. Specifically, we discuss the perihelion precession  and the bending angle of light for these two different spherically symmetric spacetimes derived in references \cite{Maeda:2006hj} and \cite{Sotiriou:2014pfa} respectively. The later in particular, applies only to the black hole spacetimes. We further delineate on the numerical bounds of relevant parameters of these theories from such computations.
\end{abstract}
\section{Introduction}\label{BY_Intro}
\noindent
The $\Lambda$ cold dark matter ($\Lambda{\rm CDM}$) model is the simplest and so far is the most successful model of our universe. Its spectrum of successes vary from the high redshift Supernovae, the cosmic microwave background, to the structure formation and so on (see, e.g., \cite{Weinberg:2008zzc} and also references therein). Nevertheless, apart from its most overwhelming  successes, there is no  explanation of the observed tiny value of $\Lambda \sim 10^{-52} {\rm m^{-2}} \equiv {\cal O }(10^{-3} {\rm eV})^4$, from a particle physics perspective, so far. There have indeed been numerous attempts including the Supersymmetry to explain this by relating $\Lambda$ to the vacuum energy density, but all such attempts either have various theoretical or observational inconsistencies (see \cite{Martin:2012bt} and also references therein). Apart from this so called cosmological constant problem, there is also the coincidence problem \cite{Polyakov,Theodore1,Woodard} too, which concerns about finding a 
satisfactory quantum mechanism that 
could explain $\Lambda$'s current tiny value, starting from some initial higher value. Also, in the observational front, it has been argued in \cite{Varun} from the Baryon Acoustic Oscillation data that the Hubble expansion rate is actually smaller than what is predicted by the $\Lambda{\rm CDM}$, for high value of the redshift, $z>2$. Another relevant problem with the $\Lambda{\rm CDM}$ is that there has been no detection of any dark matter particle candidate so far.

These conceptual or observational problems pertaining dark energy and the dark matter, namely the dark universe puzzle, have led to,  in recent times, vigorous research in alternatives to the $\Lambda{\rm CDM}$. The chief agenda of these alternative models is to generate the effect of the dark energy and/or dark matter through some additional dynamical matter fields or through the modification of the Einstein-Hilbert action or even Newton's law, without invoking any $\Lambda$ or cold dark matter by hand. We refer our reader to \cite{review, Bull:2015stt} for vast reviews and exhaustive list of references in these directions. At this stage we should mention that there exists an alternative noteworthy proposal to the solution of the cosmological constant problem, by merely treating $\Lambda$ to be a fundamental constant associated with the spacetime \cite{Dadhich:2010ca}.
  
Certainly, any such alternative model is subject to observational tests based upon the existing data. One of such fronts for doing phenomenology would be the solar system, for which we have excellent data at hand, which involve the perihelion precession of the Mercury, the bending of light by the Sun and gravitational time delay \cite{Weinberg:1972kfs,Will,Paddy}. Clearly, in order to do so in the context of the alternative gravity models, one requires static, spherically symmetric solutions with gravitational potential falling off as $\sim 1/r$. We refer our reader to \cite{Sotiriou:2013qea,Babichev:2016rlq,Wiltshire:1985us,Yunes:2011we,Charmousis:2008kc,Pani:2011xm} and references therein for an account of black hole and star solutions in scalar-tensor theories of gravity. All such theories contain additional parameters owing to the modification/extension of the action associated with the $\Lambda{\rm CDM}$ model. Therefore, the chief agenda of constraining such alternative theories are basically to 
constrain those additional parameters. For example, the solar system constraint for the Brans-Dicke theory \cite{Brans}, can be seen in \cite{review}. For the galileons (e.g. \cite{Martin-Moruno:2015kaa}), which are viable alternatives to the dark energy, such constraints can be found in \cite{Tessore:2015sma,Iorio:2012pve,Andrews:2013qva}. A further subclass of Horndeski theories, known as $f(R)$ gravity was introduced in \cite{Starobinsky} and latter has been used extensively in cosmological (see \cite{Nojiri:2010wj,Sotiriou:2008rp} for a review) and extra dimensional scenarios \cite{Chakraborty:2016ydo,Chakraborty:2014zya,Chakraborty:2014xla}. We refer our reader to \cite{Berry,Capozziello:2007ms,Capozziello:2006jj} for discussions on the solar system physics and Newtonian limits of $f(R)$ gravity. We further refer our reader to~\cite{Iorio:2012cm} for a discussion on constraining $f(T)$ gravity models and to~\cite{Iorio:2016sqy} for  constraining the Schwarzschild-de Sitter spacetime obtained in modified 
gravity theories.

In this paper, we shall work out perihelion precession and bending of light by a central massive object (which could be a black hole or the Sun) in the two static and spherically symmetric solutions, presented in \cite{Maeda:2006hj,Sotiriou:2014pfa}.
Quite interestingly, the later solution is only valid for a black hole spacetime (which, being hairy, differs from the Schwarzschild solution) and as a consequence putting forward an intriguing qualitative departure from the General Relativity, which we shall discuss in detail later. We shall further put constraints on the parameters for of the first from Mercury's perihelion precession data, which seem to match with Einstein's theory with about  $98\%$ accuracy \cite{Will}. 

The metric for the first one found for the Einstein-Gauss-Bonnet gravity, which we call the Maeda-Dadhich solution, in $n$-spacetime dimensions reads 
\begin{align}
ds^{2}=-f(r)dt^{2}+\frac{dr^{2}}{f(r)}+r^{2}d\Omega^{2}_{2}+\gamma _{AB}dx^{A}dx^{B}~,
\label{ss1}
\end{align}
with
\begin{align}
f(r)=1+\frac{r^2}{2(n-4)\alpha}\left[1\mp \left(1- \frac{2(n-5) (2n-11)-\Theta}{6(n-5)^2} + \frac{4(n-4)^2 \alpha^{\frac32} m}{r^3} -  \frac{4(n-4)^2 \alpha^2 q}{r^4}\right)^{\frac12} \right]~,
\label{ss0}
\end{align}
where $\alpha$ is the Gauss-Bonnet coupling parameter, $m$ and $q<0$  are arbitrary dimensionless constants and $A,B$ runs over the extra $(n-4)$ spacetime dimensions. A very interesting property of this solutions is, that it does have {\it no} general relativistic limit, $\alpha \to 0$. The asymptotically flat solution is found when we take $\Theta=2(n-5) (2n-11)$ and the `minus' sign. However this is a strong assumption on the extra dimensional Einstein space with metric $\gamma _{AB}$, since $\Theta$ is the Weyl tensor squared of the Einstein space \cite{Maeda:2006hj}. Even though there is no demonstration that such Einstein spaces do not exist, to be on firm ground we shall leave $\Theta$ arbitrary. This results in inclusion of the asymptotic (anti-)de Sitter branch as well. However, since the cosmological constant is `tiny', the (anti-)de Sitter term would have practically no observable effect on the solar system physics and hence we may safely ignore it for our current purpose. Thus  the relevant 
metric function reads
\begin{align}
f(r)\approx 1- \frac{(n-4)\sqrt{\alpha} m}{r} +\frac{(n-4) \alpha q} {r^2} +{\cal O}(r^{-4})~.
\label{ss2}
\end{align}
We make the identification in terms of the usual ADM mass and the Newton's constant, $(n-4)\sqrt{\alpha} m\equiv 2GM$, so that we have
\begin{align}
f(r)\approx 1- \frac{2GM}{r} +\frac{4 G^2 M^2\widetilde {q}}{r^2}~,
\label{ss1'}
\end{align}
where $\widetilde {q}= \frac{q}{m^2 (n-4)}$ is a dimensionless parameter\footnote{In this context we would like to mention that $q$ (and hence $\widetilde{q}$) must be negative in order for it to be in resonance with mass of the black hole under consideration, unlike the electric charge in \RN metric. This peculiar feature arises since it being a manifestation of bulk spacetime curvature is gravitational in character \cite{Maeda:2006hj}.}. We shall attempt to constrain this particular parameter in the following, by working on the $t, r, \varphi$ plane, where $\varphi$ is the azimuthal coordinate and the other angular coordinate will be set to $\pi/2$, while {\it all} the extra dimensional coordinates will be set to some constant values. 

Note that the above solution is not valid in the usual four spacetime dimensions, since the Gauss-Bonnet invariant becomes a total divergence in this case and hence the Maeda-Dadhich solution is essentially a higher dimensional one. However, in the usual scenarios with extra dimensions (such as the brane-world), the matter fields (for example, the light ray, Mercury or in general any Standard Model field) are essentially confined to four spacetime dimensions \cite{Maartens:2003tw}. One could consider the backreaction of such matter fields upon the extra dimensions (such as the backreaction due to dark matter/dark energy), but they seem to be relevant in much higher length scales pertaining the cosmology, than the solar system. In other words, we take the above solution to be a reasonable model in dealing with the small scale solar system physics, while still incorporating the effect of the extra dimensions. Keeping this in mind, and the spherical symmetry of the problem justifies working on the 
aforementioned $3$-surface in such a higher dimensional spacetime. 

One should also note that the Maeda-Dadhich solution is essentially a vacuum solution in higher spacetime dimensions. Thus to make all standard model matter fields (in particular, photons) effectively confined within the four dimensional hypersurface (brane) one can either explicitly incorporate delta function at the location of the brane in the matter action, or may require the extra dimensions to be compactified with rather small radii. Compactification in turn could make the Weyl term $\Theta$, appearing in \ref{ss0} large. In other words, for the Maeda-Dadhich solution to be consistent with the known phenomenology (e.g., smallness of the cosmological constant), some kind of fine tuning seems to be necessary. One plausible way out would be to tune the value of the parameter $\alpha$, so that one can appear at \ref{ss1'} from \ref{ss0} at small scales in order to explain the solar system physics (note that in \ref{ss1'}, we have absorbed the parameter $\alpha$ in $2GM$). However, it would 
be interesting to check whether such a prescription yields consistent results at a cosmological scale, since then one has to incorporate time dependent matter (e.g., dark matter, dark energy) in the setup. We hope to come back to this issue in a future work.

The other solution we shall be concerned with, is a solution with scalar field $\phi$, linearly coupled with the Gauss-Bonnet invariant ${\cal G}$, ($\equiv R_{abcd}R^{abcd} -4R_{ab}R^{ab}+R^2$) via the coupling constant $\alpha$ \cite{Sotiriou:2014pfa}, which we call the Sotiriou-Zhou solution. The relevant action reads 
\begin{align}
S= \frac{1}{16\pi G} \int d^4 x \sqrt{-g} \left[R-(\nabla_a \phi)(\nabla^a \phi)+2\alpha \phi {\cal G}\right]~.
\label{ss3'}
\end{align}
The solution was originally found in the context of the violation of the black hole no hair theorem. The metric and the configuration of the scalar field at large radial distances are given by,   
\begin{align}
ds^{2}=-f(r)dt^{2}+ h(r) dr^{2}+r^{2}d\Omega^{2}~,
\label{ss3}
\end{align}
where
\begin{align}
f(r)&\approx 1-\frac{2GM}{r}+\frac{GM P^2}{6r^3}+{\cal O}(r^{-4}), \quad h(r) \approx 1+\frac{2GM}{r}+\frac{8 G^2M^2- P^2}{2 r^2}+{\cal O}(r^{-3}),
\nonumber
\\
\phi(r)& \approx \frac{P}{r} +\frac{GMP}{r^2}+{\cal O}(r^{-3})~.
\label{ss4}
\end{align}
Here the constant $P$ (having dimension of length) can be thought of as the charge associated with the scalar field. We emphasize here that while the Maeda-Dadhich solution is equally applicable to black holes and stars, the above one is valid {\it only for black holes} (for star solution, we must have $P=0$, reducing~\ref{ss4} to the Schwarzschild spacetime).   This originates from the fact that the integral of the Gauss-Bonnet invariant over a simply connected region vanishes identically in four  spacetime dimensions leading to a trivial solution ($\phi=0$) for the scalar field, see e.g.~\cite{Berti:2015itd} and also references therein. In fact it is easy to relate the scalar charge $P$ of a black hole spacetime to the integral of the Gauss-Bonnet invariant as follows. The equation of the motion of the scalar is obtained from~\ref{ss3'}
\begin{align}
\Box \phi+\alpha {\cal G}=0
\label{ad1}
\end{align}
In the static spacetime of~\ref{ss3}, it is possible to project the above equation in the $(r,\theta,\phi)$ family of hypersurfaces (see, e.g. \cite{Bhattacharya:2007zzb})
\begin{align}
D_{a}\left(\sqrt{f}D^{a}\phi \right)+\sqrt{f}\alpha {\cal G}=0
\label{ad2}
\end{align}
where $D_a$ is the covariant derivative operator on the $(r,\theta, \phi)$ hypersurfaces. Integrating the above equation over the hypersurface one obtains two surface integrals over 2-spheres, on the horizon and at infinity respectively. The horizon integral vanishes since $f=0$ (or $h^{-1}=0$) there and also the scalar field and its derivatives are bounded, necessary for a non-singular spacetime. Putting in the asymptotic boundary conditions: $f(r), h(r) \sim 1+{\cal O}(1/r)$ and $\phi \sim P/r$ at spatial infinity, we obtain a formal expression for the scalar charge, $P$ in terms of the Gauss-Bonnet invariant as,
\begin{align}
P=\frac{\alpha}{4\pi}\int \sqrt{h}r^{2}\sin \theta drd\theta d\phi \sqrt{f}{\cal G}
\label{ad3}
\end{align}
We also note here that the first order contribution to the perihelion precession comes from the ${\cal O}(r^{-2})$ term from the expansion of $f^{-1}(r)$ and from the ${\cal O}(r^{-1})$ term of $h(r)$ \cite{Weinberg:1972kfs}. Thus it is obvious that the above solution,~\ref{ss4}, would give rise to effects identical to that of the General Relativity even for black hole spacetimes, at the leading order. Nevertheless, we would give expressions that would correspond to higher order corrections. More importantly, we would next include a coupling of the scalar field with the geodesic and would constrain the parameter, say $\beta$, corresponding to that coupling via the effective metric $\widetilde{g}_{ab}=e^{2\beta \phi} g_{ab}$. This would indeed give, as we shall see, departure from the General Relativity in the leading order. Such couplings are common in scalar-tensor or Horndeski gravity theories admitting no non-minimal coupling of the scalar with the Ricci scalar, see e.g. 
\cite{Iorio:2012pve, Andrews:2013qva, Bhattacharya:2015chc} in the context of galileons and can always be motivated from the Brans-Dicke theory in the Einstein frame. Here we also refer our reader to~\cite{Tretyakova:2016knb}, where some pathologies of geodesic motion has been reported, for a solution belonging to a class of Horndeski model different from what we consider here. 

The rest of the paper is organized is as follows. In the next section we briefly outline the basic geometric techniques. \ref{Peri} and \ref{Bend} respectively concern with the perihelion precession and the light bending calculations. Our analysis will chiefly be based on the formalism developed in \cite{Weinberg:1972kfs,Will,Paddy}. Finally, we conclude in \ref{Conc}. We shall set $c=1$ throughout. 
\section{The basic formalism and the key equations}
\noindent
We begin by considering a general ansatz for a static and spherically symmetric metric describing a black hole or star, 
\begin{align}
ds^2=-f(r)dt^2+h(r) dr^2 +r^2 d\Omega^2~.
\end{align}
Since the metric is spherical we set $\theta=\pi/2$ and consider a geodesic $u^a=\frac{dx^a}{d\lambda}$ ($u^a\nabla_au^b=0$) moving on that plane. For the higher dimensional Maeda-Dadhich solution in \ref{ss1}, we set the polar coordinate to $\pi/2$, exercising the spherical symmetry, along with setting extra dimensional coordinates to constant values. The spacetime admits two Killing vector fields on that plane -- the timelike Killing vector field $(\partial_t)^a$ and the azimuthal Killing vector field $(\partial_{\varphi})^a$, giving rise to the conserved energy ($E$) and the orbital angular momentum $(L)$ : 
\begin{align}
E=-u^a(\partial_t)_a= f(r)\frac{dt}{d\lambda}; \quad L=   u^a(\partial_{\varphi})_a = r^2\frac{d\varphi}{d\lambda}~,
\label{ss5}
\end{align}
where $\lambda$ is any affine parameter along the geodesic. The norm of the geodesic is $u^au_a=-\kappa$, where $\kappa$ equals $1(0)$ for timelike(null) geodesics. Putting these all in together, we get
\begin{align}
\frac{dr}{d\lambda} = h^{-\frac12}(r)\left[E^2/f(r) - (L^2/r^2+\kappa) \right]^{\frac12}~,
\label{ss6}
\end{align}
which, when combined with the second of \ref{ss5} gives
\begin{align}
\frac{d\varphi}{d r} = \frac{  h^{\frac12}(r)/r^2 }{\left[E^2/(L^2f(r)) - (1/r^2+\kappa/L^2) \right]^{\frac12}}~,
\label{ss7}
\end{align}
which is the key equation we shall use to deal with both bound (pertaining the perihelion precession of a massive test object around a black hole or star, $\kappa=1$) and unbound (i.e., for the deflection of light, with $\kappa=0$) orbits.

As we mentioned in the introduction, for the Sotiriou-Zhou solution (see \ref{ss3'} to \ref{ss4}), we shall also consider coupling of the scalar field with the test particle. Let us now take an account of the modification caused by this. The test particle couples with the field $\phi$ via the `effective' metric  $\widetilde{g}_{ab}=e^{2\beta \phi} g_{ab}$, inspired from the Brans-Dicke theory in the Einstein frame \cite{Iorio:2012pve,Andrews:2013qva,Bhattacharya:2015chc}. The action for a test particle reads in that case
\begin{eqnarray}
S=-\int \sqrt {-\widetilde{g}_{ab}dx^a dx^b}=-\int d\lambda e^{\beta\phi}  \sqrt {-g_{ab}\frac{dx^a}{d\lambda}\frac{ dx^b}{d\lambda} }~.
\label{g23}
\end{eqnarray}
We get the equation of motion from the variation of the above action,
\begin{eqnarray}
u^a\nabla_a u_b= 2\beta (u\cdot u) \nabla_b\phi~.
\label{g24}
\end{eqnarray}
Clearly, unlike geodesics the norm of the tangent vector is not preserved along the trajectory for a non-constant $\phi$.  We contract the above equation by $u^b$ to find,
\begin{eqnarray}
u^a\nabla_a (u\cdot u)= 4\beta (u\cdot u) u^a \nabla_a \phi~.
\label{g25}
\end{eqnarray}
The directional derivatives acting on scalars are just partial derivatives, $u^a\nabla_a (u\cdot u)=\partial_{\lambda}(u\cdot u) $ and $u^a\nabla_a \phi= \partial_{\lambda}\phi$, so that the above equation gives the {\it non-conserved norm} of the tangent vector,
\begin{eqnarray}
u\cdot u=-\kappa e^{4\beta \phi}~,
\label{g26}
\end{eqnarray}
with $\kappa$ being an integration constant. Setting $\beta=0$ should recover the  geodesic and hence we must set $\kappa=1$ for timelike particles. For null geodesics, we must set $\kappa=0$, and hence the norm will be preserved for them. Certainly, this shows that even though the motion the null geodesics remain formally the same as earlier, it does not remain so for the timelike geodesics. In particular, the non-conservation of the norm, \ref{g26} shows qualitative departure from the geodesic motion.

Since the scalar field is independent of $t$ and $\varphi$, \ref{g24} shows that similar to the geodesic, we would have conserved quantities of~\ref{ss5}. Then using \ref{g26} it is straightforward to get a modification of \ref{ss7},   
\begin{align}
\frac{d\varphi}{d r} = \frac{  h^{\frac12}(r)/r^2 }{\left[E^2/(L^2f(r)) - (1/r^2+\kappa e^{4\beta \phi}/L^2) \right]^{\frac12}}~.
\label{ss8}
\end{align}
With these necessary equipment, we are now ready to go into the explicit computations for the perihelion precession and bending of light.
\section{The perihelion precession}\label{Peri}

\subsection{The Maeda-Dadhich solution}
\noindent
For the Maeda-Dadhich solution presented in \ref{ss1}, we have $h=f^{-1}$ in \ref{ss7}. Let us denote for a closed orbit, the position of aphelion and perihelion by $r=r_{+}$ and $r=r_{-}$ respectively. Then we must have $\frac{dr}{d\lambda}\big\vert_{r=r_{\pm}}=0$, which implies the denominator of \ref{ss7} to be vanishing there (with $\kappa=1$). Using this, we eliminate the two constants of motion $E$ and $L$ in favour of these two radii,
\begin{align}
E^2=\frac{r_{+}^{2}-r_{-}^{2}} {\frac{r_{+}^{2}}{f(r_{+})}-\frac{r_{-}^{2}}{f(r_{-})}},\quad
L^2=\frac{{f(r_{+})}-{f(r_{-})}}{\frac{f(r_{+}) }{r_{+}^{2}}-\frac{f(r_{-})}{r_{-}^{2}}}~.
\label{ss9}
\end{align}
We shall consider orbits whose radii are much larger than the Schwarzschild radius of the central mass and hence will  make a week field expansion of the metric. Retaining only up to quadratic order  terms in the expansion of $f^{-1}(r)$, it is clear that we could write the argument of the square root of the denominator of \ref{ss7} as 
\begin{align}
\left[E^2/(L^2f(r)) - (1/r^2+1/L^2) \right]= C \left(\frac{1}{r_{-}}-\frac{1}{r}\right)\left(\frac{1}{r}-\frac{1}{r_{+}}\right)~,
\label{ss10}
\end{align}
where the constant $C$ can be determined by letting $r\to \infty$, $C=r_+ r_-(1-E^2)/L^2$. Taking this limit, using~\ref{ss9} and the explicit expression for $f(r)$ from \ref{ss1'},  we get
\begin{align}
C=\frac{2GM(r_{-}-r_{+}) }{2GM (r_{-}-r_{+})+ 4G^2M^2(1-\widetilde{q}) \left(\frac{r_{-}}{r_{+}}-\frac{r_{+}}{r_{-}}\right)} \approx
1-2GM (1-\widetilde{q}) \left(\frac{1}{r_{+}}+\frac{1}{r_{-}}\right)~,
\label{ss11}
\end{align}
where in the last equality we have once again used the fact that we are working in a weak field regime, $r_{\pm}\gg 2GM$. Substituting \ref{ss10} and \ref{ss11} into \ref{ss7}, we get after integration
\begin{align}
\varphi (r)-\varphi (r_{-})\approx \left(1+GM (1-\widetilde{q}) \left(\frac{1}{r_{+}}+\frac{1}{r_{-}}\right)\right)  \int _{r_{-}}^{r}dr \frac{(1+GM/r)}{r^2\sqrt{\left(\frac{1}{r_{-}}-\frac{1}{r}\right)\left(\frac{1}{r}-\frac{1}{r_{+}}\right)}}~.
\label{ss12}
\end{align}
The above integration can be easily performed using a new angular variable $\psi$ \cite{Weinberg:1972kfs},
\begin{align}
\frac{1}{r}=\frac{1}{2}\left(\frac{1}{r_{+}}+\frac{1}{r_{-}}\right)+\frac{1}{2}\left(\frac{1}{r_{+}}-\frac{1}{r_{-}}\right)\sin \psi ~,
\label{int}
\end{align}
so that $\psi=\mp \pi/2$ correspond to the perihelion (aphelion), respectively. 

For our purpose, we need to obtain the contribution of the full closed orbit, starting from the perihelion and coming back to it, via the aphelion. Due to the symmetry of the problem, it is clear that the contribution to the above integral originating due to the path from $r_-$ to $r_+$ would exactly be the same as that from $r_+$ to $r_-$. Then the precession angle of the perihelion is clearly given by $\delta \varphi =2\left[\varphi (r_{+})-\varphi (r_{-})\right]-2\pi$. Evaluating the integral in \ref{ss12} using~\ref{int}, we find
\begin{align}
\delta\varphi =  3\pi GM (1-2\widetilde{q}/3) \left(\frac{1}{r_{+}}+\frac{1}{r_{-}}\right)~,
\label{ss13}
\end{align}
where the Newtonian limit corresponds to $M/r_{\pm} \to 0$ and the General Relativistic limit corresponds to $\widetilde{q}=0$ \cite{Weinberg:1972kfs}. Since $\widetilde{q}<0$, the precession angle actually increases with increase of $|\widetilde{q}|$.  We may ascribe this behaviour of the precession angle to the gravitational nature of $|\widetilde{q}|$, as mentioned earlier. This is in contrast with the case of an electric charge~\cite{Chakraborty:2012sd}.

We shall now apply the above expression for the perihelion precession in the context of Mercury.
The observed value of Mercury's perihelion precession is about $43.03$ arc seconds per century \cite{Will}. Now, it constrains any factor multiplying the pure General Relativity term in \ref{ss13} to be $1.003\pm 0.005$. This yields
\begin{align}
|\widetilde{q}| < 0.024,
\end{align}
which is consistent with earlier results providing bounds on the Gauss-Bonnet parameter $\alpha$ \cite{Chakraborty:2012sd,Chakraborty:2013ywa,Chakraborty:2015vla}. Hence the charge parameter $\widetilde{q}$ has a pretty small value which is quiet expected. Thus we have performed the first task in relation to Maeda-Dadhich solution. We will now consider perihelion precession for the Sotiriou-Zhou solution before returning to the constraints on $\widetilde{q}$ from bending angle of light in the next section.
\subsection{The Sotiriou-Zhou solution}\label{sg}
\noindent
We shall now deal with the Sotiriou-Zhou solution (see \ref{ss4}). For this, first we shall do our computations by setting $\beta \neq 0$ in \ref{ss8} (with $\kappa=1$) and then we shall consider the situation when $\beta=0$. As we mentioned in Sec.~1, since the ${\cal O}(r^{-2})$ term in the expansion of $f^{-1}(r)$ in the latter case is identical to that of General Relativity, and consequently, we shall have no leading departure from the General Relativity for  $\beta =0$. Nevertheless, we shall briefly outline below the higher order computation as well.  

Concentrating on the case $\beta\neq 0$, $(dr/d\lambda)\vert_{r_{\pm}}=0$ as earlier, implies the denominator of \ref{ss8} to be vanishing there. Using this, we now find an analogue of \ref{ss9},
\begin{align}
E^2&=\left(r_{+}^{2}e^{4\beta \phi(r_+)}-r_{-}^{2}e^{4\beta \phi(r_-)}\right)\left(\frac{r_{+}^{2}}{f(r_{+})}-\frac{r_{-}^{2}}{f(r_{-})}\right)^{-1}
\nonumber
\\
L^2&=\left({f(r_{+}) e^{4\beta \phi(r_+)}  }-{f(r_{-}) e^{4\beta \phi(r_-)}}\right)\left(\frac{f(r_{+}) }{r_{+}^{2}}-\frac{f(r_{-})}{r_{-}^{2}}\right)^{-1}~.
\label{ss14}
\end{align}
We now expand the function $\exp(4\beta \phi)$ using \ref{ss4}, up to terms of ${\cal O}(r^{-2})$ and linear in the coupling constant $\beta$, $\exp(4\beta \phi)\approx 1+(4\beta P/r)+(4\beta GMP/r^2)$. Likewise, we keep only up to ${\cal O} (r^{-2})$ terms in the expansion of $f^{-1}(r)$, to write the denominator of \ref{ss8} as 
\begin{align}
E^2/(L^2f(r)) - (1/r^2+\kappa e^{4\beta \phi}/L^2) \approx C \left(\frac{1}{r_{-}}-\frac{1}{r}\right)\left(\frac{1}{r}-\frac{1}{r_{+}}\right)~.
\label{ss15}
\end{align}
Following similar steps as earlier, we obtain
\begin{align}
C\approx
1-2GM (1-\beta P/GM) \left(\frac{1}{r_{+}}+\frac{1}{r_{-}}\right)~,
\label{ss16}
\end{align}
which yields the precession angle
\begin{align}
\delta\varphi \approx  3\pi GM (1-3\beta P/GM) \left(\frac{1}{r_{+}}+\frac{1}{r_{-}}\right)~.
\label{ss17}
\end{align}
Similar arguments as earlier leads to a constraint $|\beta P/GM| < 0.003 $. Note that since the scalar charge has dimension of length, $P/GM$ is a dimensionless quantity and thus could be regarded as the scalar charge per unit mass or specific scalar charge. Thus the dimensionless quantity we have constrained is nothing but the specific scalar charge scaled by the matter-scalar coupling constant. Further note that depending on sign of the combination $\beta P$, the precession angle either increases or decreases. For $\beta P<0$ the precession angle increases, while for $\beta P>0$ it decreases. We emphasize here that  the coupling parameter $\beta$ only appears in the combination $\beta P$, thus in the case of vanishing scalar hair (i.e., $P=0$, such as in the Solar system) the effect of $\beta$ would no longer be present. Certainly, this seems to be a potentially interesting phenomenon distinguishing between the black hole and the non-black hole structures, absent in the General Relativity.

Finally, as promised we will mention the perihelion precession angle for the case $\beta =0$ by retaining higher order terms. Following similar arguments as earlier, we obtain in the place of \ref{ss12},
\begin{align}
\phi (r)-\phi (r_{-})=\frac{1}{\sqrt{C}}\int _{r_{-}}^{r}dr \frac{1}{r^{2}}\frac{\sqrt{h(r)}}{\sqrt{\left(\frac{1}{r_{-}}-\frac{1}{r}\right)\left(\frac{1}{r}-\frac{1}{r_{+}}\right)}}~,
\end{align}
where $C$ is given by \ref{ss11}. Indeed, expanding both $\sqrt{h(r)}$ and $f^{-1/2}(r)$ to ${\cal O}(r^{-2})$, we find, following the technique described earlier, the expression for the perihelion precession angle, 
\begin{align}
\Delta \varphi =3\pi GM\left(\frac{1}{r_{+}}+\frac{1}{r_{-}}\right)+\frac{\pi G^{2}M^{2}}{4}\left(1-\frac{P^2}{2G^2M^2}\right )\left(\frac{1}{r_{+}}+\frac{1}{r_{-}}\right)^{2}~.
\end{align}
However in order to derive the above result we have expanded the metric functions to $\mathcal{O}(r^{-2})$ only. Since the correction to the metric component $f(r)$, due to existence of scalar charge, comes only at the order $\mathcal{O}(r^{-3})$, for a  consistent analysis one should expand the metric functions to $\mathcal{O}(r^{-3})$. When that is done the precession angle gets modified and receives a contribution from $GMP^{2}$, such that,
\begin{align}
\Delta \varphi =3\pi GM\left\lbrace 1-\frac{P^{2}}{4}\left(\frac{1}{r_{+}}+\frac{1}{r_{-}}\right)^{2} \right\rbrace \left(\frac{1}{r_{+}}+\frac{1}{r_{-}}\right)+\frac{\pi G^{2}M^{2}}{4}\left(1-\frac{P^2}{2G^2M^2}\right )\left(\frac{1}{r_{+}}+\frac{1}{r_{-}}\right)^{2}~.
\end{align}
At this stage one can easily check that the $\mathcal{O}(MP^{2})$ term is suppressed by a factor $(GM/c^{2})\times R^{-1}$, with $R \sim {\cal O} (r_{\pm})$. Further as one computes the precession angle it turns out that $\mathcal{O}(M^{2})$ term is also five orders of magnitude smaller compared to the $\mathcal{O}(P^{2})$ term. Another important point to note is the following fact --- the precession angle always decreases with increase of the scalar charge $P$ independent of its sign.

As we have discussed earlier, the Sotiriou-Zhou solution with $P\neq 0$ is applicable only for black holes and accordingly, we cannot use the above expressions in the context of the Solar system.  However, we point out here a golden window where we may apply our results, pertaining directly to a black hole spacetime. Rather recently, in~\cite{Schodel:2002vg}, the observational evidence of a supermassive black hole (namely, the Sgr A* with mass $M=(3.7\pm 1.5) \times  10^{6}M_{\odot}$) at the centre of the Milky Way was obtained,  by observing the orbital motion of a star for around ten years, called S2. The star has an orbital period of  $15.2$ years and the pericentre distance from  Sgr A* is  $17$ light hours. More recently, it was proposed in~\cite{Slovak} to 
actually observationally measure the precession of the perihelion of S2 around Sgr A* and to compare with theoretical predictions which is about $66$ times larger than that of the Mercury.  However, the complete data regarding this is yet to come out. Since the Sotiriou-Zhou solution clearly distinguishes between black holes and other compact objects, the above arena would be an excellent testbed for various alternative gravity models and hairy black holes. We sincerely do hope that when in near future the full data comes out, the expressions we have derived would be very useful and illuminating. In the context of strong gravitational lensing and shadow formation for the same black hole, we further refer our reader to~\cite{Abdujabbarov:2015xqa} and references therein.  
\section{The bending of light}\label{Bend}
\noindent
We shall present below computations on the bending of light in the two backgrounds we are concerned with. Our analysis will chiefly be based on the formalism described in \cite{Paddy}. Alike perihelion precession we will attempt to constrain any parameters of the theories considered here as well. However there are two difficulties that one encounters, firstly, the data related to this is not as accurate as Mercury's perihelion precession \cite{Weinberg:1972kfs,Will}. Secondly, it would turn 
out that the corrections to the bending of light due to these alternative gravity models comes at higher order computations. Nevertheless, we shall provide expressions of the bending angle for both these theories, below, besides obtaining constraints on the parameters of the theories.
\subsection{The Maeda-Dadhich solution}
\noindent
Let us start by considering bending angle of light in connection to Maeda-Dadhich solution. Since light ray propagates along the null geodesic, at least in the weak gravity regime, we set $\kappa=0$ in \ref{ss6}, \ref{ss7} and \ref{ss8}. Defining a new variable $r=1/u$ and the impact parameter $J=L/E$, we get from \ref{ss7} for the Maeda-Dadhich solution \ref{ss1} the following,  
\begin{align}
\left(\frac{du}{d\varphi}\right)^{2}=\frac{1}{J^{2}}-u^{2}\left(1-2MGu+4G^2M^2 \widetilde{q}u^{2}\right)~.
\label{lb1}
\end{align}
Differentiating again with respect to $\varphi$ we obtain, 
\begin{align}
\frac{d^{2}u}{d\varphi ^{2}}+u={3MG}u^{2}-8M^2G^2\widetilde{q}u^{3}~.
\end{align}
We now make the ansatz, $u=\frac{1}{J}\cos \varphi +v(\varphi)$ with $Jv\ll 1$, which expectedly implies that the correction term $v$ is much less than the impact parameter $1/J$. This yields  
\begin{align}
\frac{d^{2}v}{d\varphi ^{2}}+v&=\frac{3 MG}{J^{2}}\cos ^{2}\varphi-\frac{8M^2G^2\widetilde{q}}{J^{3}}\cos^{3}\varphi
=\frac{3MG}{2J^{2}}\left(1+\cos 2\varphi \right)-\frac{2M^2 G^2\widetilde{q} }{J^{3}}\left(3\cos \varphi+\cos 3\varphi \right)~.
\end{align}
In order to simplify further, let us take $v=(3MG)/(2J^{2})+\chi (\varphi)$, giving
\begin{align}
\frac{d^{2}\chi}{d\varphi^{2}}+\chi=\frac{3MG}{2J^{2}}\cos 2\varphi -\frac{ 2M^2G^2 \widetilde{q}}{J^{3}}\left(3\cos \varphi+\cos 3\varphi \right)~.
\end{align}
The above equation can easily be solved for the function $\chi(\varphi)$, yielding the following solution for the inverse radial function $u(\phi)$,
\begin{align}
u(\varphi)= \frac{1}{J}\left(1-\frac{9 M^2G^2 \widetilde{q}}{4J^{2}}\right)\cos \varphi +\frac{2MG}{J^{2}}~.
\end{align}
From which $\cos \varphi$ can be found in the asymptotic region by letting the inverse radial function $u\to 0$,
\begin{align}
\cos\varphi    \approx -{\frac{2MG}{J}}  \left(1+\frac{9 M^2G^2 \widetilde{q}}{4J^2}\right)~.
\end{align}
Certainly, the right hand side is much smaller than unity, as the impact parameter $J$ must be located outside the Sun and the actual radius of it is much larger than its Schwarzschild radius, $2MG$. Now writing $\cos\varphi=\sin(\pi/2-\varphi)$ and $\sin^{-1}x\approx x$ for $x\ll1$, we find 
\begin{align}
\varphi-\pi/2 =\frac{2MG}{J}\left(1+ \frac{9 M^2G^2 \widetilde{q}}{4J^2}  \right)~.
\end{align}
The deficit in the azimuthal angle is simply the twice of the above, owing simply to the symmetry of the problem. In other words,
the deficit angle for light for the Maeda-Dadhich spacetime \ref{ss1} becomes
\begin{align}
\delta\varphi =\frac{4MG}{J}\left(1+ \frac{9 M^2G^2 \widetilde{q}}{4J^2}  \right)~.
\label{bend'}
\end{align}
Certainly, the correction due to the dimensionless parameter $\widetilde{q}$ is of higher order in $MG/J$, compared to that of the leading order. The effect of the parameter $\widetilde{q}$ is to decrease the bending angle since it is a negative quantity. However this effect is exactly opposite to that presented earlier in the context of perihelion precession, where precession angle increases with $\widetilde{q}$. Further the observed value of bending angle of light from Sun corresponds to $1.75(1\pm 0.02)~\textrm{arc second}$~\cite{Will}. From the \gr\ expression one can immediately read off the numerical value of $MG/J$, which when substituted into the correction term of~\ref{bend'}, leads to $|\widetilde{q}|<0.046$. Thus it is consistent with the constraint $|\widetilde{q}|<0.024$ obtained from Mercury's perihelion precession. The bound corresponding to the perihelion precession is more improved, just because the data available for it is much finer than the light bending~\cite{Will}. 
\subsection{Sotiriou-Zhou solution}
\noindent
Let us finally consider the Sotiriou-Zhou solution, as presented in \ref{ss3} and \ref{ss4}. The analogue of \ref{lb1} now reads
\begin{align}
\left(\frac{du}{d\varphi}\right)^{2}=\frac{1}{J^{2}}-u^2\left(1-2MG u -P^2/2J^2\right)~,
\end{align}
which, upon differentiation yields the second order differential equation for the variable $u=1/r$ in powers of $u$,
\begin{align}
\frac{d^{2}u}{d\varphi ^{2}}+\left(1-P^2/J^{2}\right)u={3GM}u^{2}~.
\end{align}
Let us now employ a suitable ansatz $u=(1/J)\cos \omega\varphi +v(\varphi)$, where $\omega^{2} =1-P^2/J^{2}$ and $Jv\ll1$.  Using the above ansatz we finally obtain the following differential equation for $v(\varphi)$,
\begin{align}
\frac{d^{2}v}{d\varphi^{2}}+\left(1-P^2/J^2\right)v=\frac{3GM}{J^{2}}\cos^2 \omega \varphi~.
\end{align}
The above equation can be solved as earlier and we get the deflection angle after a little algebra,
\begin{align}
\delta \varphi \approx \frac{4GM}{J}\left(1+\frac{P^{2}}{8J^{2}}\right)~.
\end{align}
Note that the correction to the deflection angle due to the scalar charge is always positive and it increases with the increase of $|P|$. This is also opposite to what we had obtained in the case of the perihelion precession, where the precession angle decreases with increase of the scalar charge $|P|$. This is also opposite to the Maeda-Dadhich solution, since there the bending angle decreases with increase of the respective parameter. Even though the Sotiriou-Zhou solution is applicable for black hole spacetimes only, one can put some bound on it by naively applying the solar system data for a solar mass black hole, leading to, $|P/GM|<0.09$. However the above is one illustrative example, in principle in the context of Sotiriou-Zhou solution, one must use the available data pertaining the week lensing for a black hole like Sgr A*, as discussed in~\ref{sg}. As emphasized earlier, this is a golden window to be opened up in the near future and then one can provide more stringent bounds on the 
scalar charge using gravitational lensing by supermassive black holes, one example being that of Sgr A* and can have a better insight on these dark energy motivated gravity models (see for example \cite{Bozza:2012by,BinNun:2010ty,Broderick:2009ph,Ghez:2005dd}).
\section{Conclusions}\label{Conc}
\noindent
In this work, we have considered Gauss-Bonnet gravity theories belonging to Horndeski models of gravity, which have their roots in attempting to explain the accelerated cosmological expansion of the universe. Considering two such variants, we have explored the influence of modifications beyond Einstein gravity on both the perihelion shift and light bending.  For the Maeda-Dadhich solution in the context of the Einstein-Gauss-Bonnet gravity, our work concerns with the possibility of having higher curvature terms' effects in the solar system phenomenology. A notable  feature corresponding to this solution is that, the parameter $\widetilde{q}$ is of gravitational origin and hence is always negative. This results in the increase in the precession angle whereas decrease in the bending angle, in contrast with the effect of the electric charge. From the present available data for these solar system phenomena, we have derived corresponding bounds on the characteristic parameter of the Maeda-Dadhich solution. 
Further, the constraint on the parameter from perihelion precession is shown to be consistent with the light bending measurement. 

Next in our analysis we have explored the role of scalar hair as demonstrated by Sotiriou and Zhou. Following  the Sotiriou-Zhou solution, we have computed the modification of the perihelion shift due to the hairy terms. Further, imposing non-minimal coupling to matter sector also changes, as we have seen the precession angle. Finally, we have considered light bending in this spacetime and have presented analytical expressions.  Contrary to the Maeda-Dadhich solution, for Sotiriou-Zhou solution, on the other hand, the precession angle decreases while the light bending angle increases as the scalar charge is increased. As we also have discussed, we cannot apply these results in the context of the Solar system, as this particular solution is not applicable for non-black hole objects. However, we have emphasized one possible window to be opened in near future pertaining the observation of the Sgr A* black hole, where our expressions could be applied directly providing possible stringent bound on the scalar hair.
   
The entire analysis we have performed above, is valid only in the weak field regime. Several interesting future avenues are hopefully to be done elsewhere. For example, the strong field version of these computations, in the context of black holes seems to be much promising and insightful. 
\section*{Acknowledgements}
\noindent
Research of S.C. is funded by a SPM fellowship from CSIR, Government of India. The authors are thankful to Naresh Dadhich and Hideki Maeda for useful comments on an earlier version of the manuscript and T. Sotiriou for pointing out some relevant references. The authors also acknowledge anonymous referees for useful comments. 

\end{document}